# Ultra-low-power Wireless Streaming Cameras

Saman Naderiparizi, Mehrdad Hessar, Vamsi Talla, Shyamnath Gollakota and Joshua R. Smith
University of Washington

## ABSTRACT

Wireless video streaming has traditionally been considered an extremely power-hungry operation. Existing approaches optimize the camera and communication modules individually to minimize their power consumption. However, the joint redesign and optimization of wireless communication as well as the camera is what that provides more power saving.

We present an ultra-low-power wireless video streaming camera. To achieve this, we present a novel "analog" video backscatter technique that feeds analog pixels from the photo-diodes directly to the backscatter hardware, thereby eliminating power consuming hardware components such as ADCs and amplifiers. We prototype our wireless camera using off-the-shelf hardware and show that our design can stream video at up to 13 FPS and can operate up to a distance of 150 feet from the access point. Our COTS prototype consumes 2.36*mW*. Finally, to demonstrate the potential of our design, we built two proof-of-concept applications: video streaming for micro-robots and security cameras for face detection.

## 1. INTRODUCTION

In this paper, we explore the following question: can we design an ultra-low-power device that can perform video streaming? A fundamental challenge in achieving such a capability is that existing low-power wireless camera systems heavily duty-cycle the camera to achieve overall lower power. For instance, RF-powered wireless camera prototypes [15, 27] require extensive duty-cycling on the order of tens of minutes, to capture, process and communicate a single frame.

To understand why this is the case, let us look at the different components in a low-resolution video-streaming device: camera and communication. Cameras have optical lens, an array of photo-diodes connected to amplifiers and finally an ADC to translate the analog pixels into digital values which are then transmitted on the wireless medium.

Existing approaches optimize the camera and communication modules individually to minimize their power consumption. However, designing a video streaming device requires power-consuming hardware components and video codec algorithms that interface the camera and the communication modules. Specifically, optical lens and photo-diode arrays can be designed to consume as little as 1.2 $\mu$W [9]. Similarly, recent work on backscatter can significantly lower the power consumption of communication to a few microwatts [10, 12], using custom ICs. However, interfacing the camera hardware with backscatter requires amplifiers, ADCs and AGCs that significantly add to the power consumption.

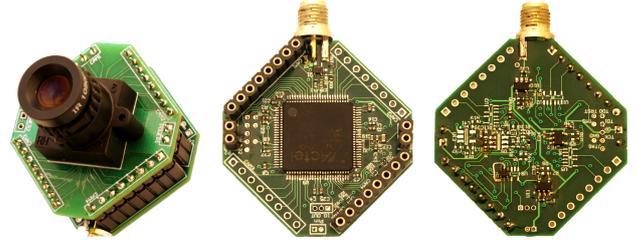

**Figure 1—Prototype of our video streaming camera.** Image of the analog camera, FPGA digital core and pulse width modulated (PWM) backscatter, all implemented using COTS components. The overall board measures 3.5 cm by 3.5 cm by 3.5 cm.

We present the design of the first ultra-low-pwoer video streaming device. We take our inspiration from Theremin's cavity resonator (popularly known as the Great Seal Bug [4]) that uses capacitive changes in a flexible metallic membrane to detect sound waves. Building on this idea, we create the first "analog" video backscatter system that does not use amplifiers, ADCs and AGCs. At a high level, we feed analog pixels from the photo-diodes directly to the backscatter hardware. We achieve this by connecting the antenna to an array of photo-diodes whose output voltage/impedance varies as a function of the pixel value; thus, eliminate power-hungry hardware components including amplifiers, AGCs and ADCs. Such an approach would have the added benefit that the video quality scales smoothly with a varying wireless channel, without the need for explicit rate adaptation.

To design such a system in practice, this paper introduces two key technical innovations:

• *We design a pulse width modulated backscatter system.* Existing amplitude modulated backscatter solutions are inefficient for streaming video. Specifically, in the absence of amplifiers, the voltage output corresponding to each pixel takes a limited range of values. Mapping these voltages directly to impedance values in the backscatter hardware results in amplitude modulated RF signals that span a limited range of amplitudes. Since the wireless channel adds noise, this results in a low SNR signal which significantly limits range. Our design instead, uses pulse width modulation (PWM) to map the pixel voltage values to different widths of a pulse. Since the access point (AP) can distinguish between different pulse widths using a higher sampling rate (4 MHz), this significantly increases our video streaming range. §3.1 describes our techniques for transforming the analog pixel values into different pulse widths using a passive ramp circuit as well as for backscattering pulse width modulated signals using sub-carrier modulation.

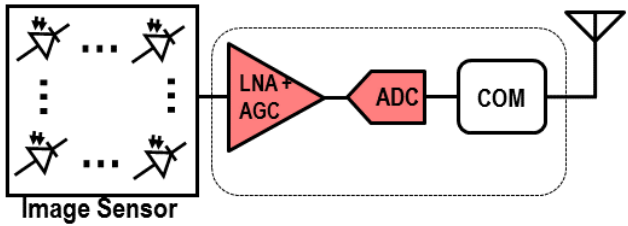

**Figure 2—Conventional Camera Design.** The amplifier, AGC, ADC and digital communication consume orders of magnitude higher power than what is available on an ultra-low-power device.

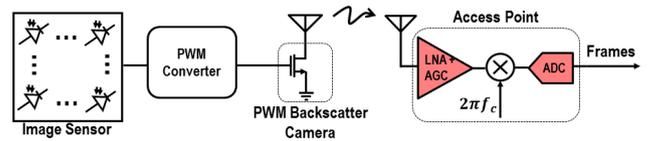

**Figure 3—Our Approach.** We eliminate the power-hungry amplifier, AGC and ADC from the camera and instead, delegate these to the access point. We use PWM backscatter to directly transmit the analog output of the camera.

• *We implemented our system using COTS components and deployed it in real scenarios.* Using a low-power FPGA, analog/mixed-signal circuits and a random pixel access camera we implemented our wireless camera system. We deployed on a small and power-constrained robot and showed that our camera can enable such a robot with video streaming capability while negligibly impacting its battery life. We also used our system in a surveillance setting and showed that more than 90% of the faces in the received video are detectable when the camera is less than 100 feet from the access point.

We implement a prototype of our backscatter design on an ultra-low power FPGA platform using a 112 × 112 grayscale random pixel access camera from CentEye [1], which provides readout access to the individual analog pixels. The access point is implemented using GNU-Radio software on a USRP-X300 hardware. We evaluate our prototype under different conditions to study its performance under different room lighting conditions, different distances from the access point and different data redundancy (i.e., coding). Our findings are summarized below:

• We can stream at 7–13 frames per second up to 150 feet from the AP, under various lighting conditions. The *PSNR* across these distances was between 24–35 dB.

• By adding redundancy to the camera video stream (reducing frame rate in exchange for coding) we can increase the *PSNR* of the video at a given distance by 3 dB.

• The power consumption of our off-the-shelf camera prototype with is about 2.36 mW.

We also develop two proof-of-concept applications: video streaming for micro-robots and security cameras in home automation. Our results show that our wireless camera design can enable a small 50 mm diameter Elisa-3 robot [2] to wirelessly stream video with negligible impact on its battery life. We also show that, for security camera and home automation applications, the quality of the wirelessly streamed video is sufficient for performing face detection.

## 2. MOTIVATION AND BACKGROUND

Fig. 2 shows the architecture of a traditional wireless camera. The output of the camera sensor (photodiodes) is first amplified by a low noise amplifier (LNA) with automatic gain control (AGC). The AGC adjusts the gain of the amplifier to ensure that the output falls within the dynamic range of the analog to digital converter (ADC). Next, the ADC (typically with 8-bit resolution) converts the analog voltage into discrete digital values. Digital data from low data rate sensors such as an accelerometer, temperature or microphone is typically transmitted by the communication block without any additional processing [6, 31]. Cameras, on the other hand, are sampled at high data rates (Mbps) and transmission of raw camera output would require a higher throughput and low latency communication link that cannot be supported by traditional wireless backscatter techniques.

Unfortunately, this architecture cannot be translated to an ultra-low-power device that operates on harvested power. Although camera sensors consisting of an array of photodiodes have been shown to operate on as low as 1.2 $\mu$W of power [9], amplifiers, AGC, ADC and the communication block require orders of magnitude higher power (10s of mW for a 112 × 112 image sensor sampled at 13 frames per second) than what is needed for their photodiode array. Furthermore, the power consumption exacerbates as we scale the resolution and/or the frame rate of the camera. Say we scale the resolution of the camera sensor by a factor of $M$ by $N$, where $M$ represents scaling for number of rows and $N$ is for the columns. The power consumption of a roller shutter based scanning photodiode sensor which is already very low (as low as a few $\mu$W) increases only by a factor of $N$. This is because in a roller shutter based scanning camera, the photodiode pixels are addressed and arranged in rows and columns. At any instance of time, only a single row of photodiodes is active and the rest are power gated to conserve power. So, the power consumption of the sensor scales with number of photodiodes in a row, i.e., by a factor of $N$, the column scaling factor. However, the sampling rate/operating frequency and the bandwidth requirements of the subsequent blocks increase by a factor of *MN* which proportionately increase the power consumption of amplifier, AGC, ADC and digital communication by a factor of *MN*. This demonstrates that these components significantly add to the power consumption as the image resolution increases. A similar analysis can also be performed for increasing the frame rate of the camera. Specifically, the power consumed by a photodiode array is independent of the frame rate whereas the power consumption of the amplifier, AGC, ADC and digital communication would increase linearly with the frame rate.

This demonstrates that the existing architectures do not scale with the resolution and/or the frame rate of the wireless camera. In particular, existing architectures use of amplifiers, AGCs and ADCs exacerbate the problem of power consumption with increased resolution and frame rate.

## 3. OUR APPROACH

Designing a continuously streaming ultra-low-power wireless camera requires us to decrease the power consumption of the wireless camera by orders of magnitude. Our key insight to reducing the power consumption of the wireless camera is to eliminate the power-hungry components — the amplifier, AGC, ADC and digital communication — from the wireless camera system. Instead, we delegate the power-hungry components to a powered fixed device in the infrastructure such as a wireless access point.

In the rest of this section, we describe how we eliminate the amplifier, AGC, ADC and digital communication on the wireless camera by using pulse width modulation backscatter.

### 3.1 Pulse Width Modulation Backscatter

*Problem.* At a high level, we eliminate the ADC, amplifier and AGC on the camera to significantly reduce the power consumption. This means that we are limited to working with analog voltage output of the photodiodes. A *naive* approach would be to leverage existing analog backscatter technique used in wireless microphones [28] and implement an analog backscatter camera. In an analog backscatter system, the output of a sensor directly controls the gate of a field-effect transistor (FET) connected to an antenna. As the output voltage of the sensor varies, it changes the impedance of the FET which amplitude modulates the RF signal backscattered by the antenna. The access point decodes the sensor information by demodulating the backscattered RF signal which is amplitude modulated. However, this approach cannot be translated to a camera sensor. The output of the photodiode has very limited dynamic range (less than 100 mV under indoor lighting conditions). These small voltage changes map to a very small subset of radar cross-sections at the antenna [7]. As a result, the antenna backscatters a very weak signal and since wireless channel and receivers add noise, this approach results in low SNR signal at the access point. This limits the system to poor signal quality and limited operating range. One can potentially get around this constraint by introducing a power-hungry amplifier and AGC but this would negate the power savings of analog backscatter.

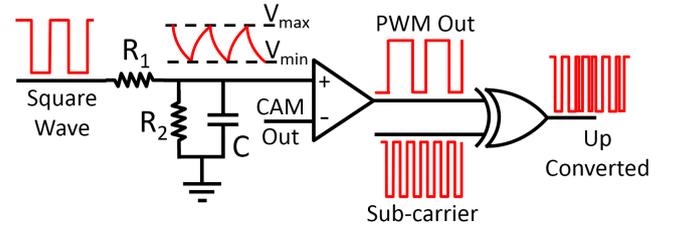

**Figure 4— Architecture of the PWM converter**.

*Our solution.* Our approach to create a high SNR backscatter signal at the antenna is to leverage digital components, which also have the added benefit of scaling in power with CMOS technology node (Moore's law). Existing backscatter communication systems such as RFID [21], Passive Wi-Fi [12], FM Backscatter [29] and Interscatter [10] all use a digital FET to modulate the radar cross section of the antenna. The digital FET is driven by a square wave which toggles the antenna between open and short impedance states resulting in maximum achievable radar cross section and consequently high SNR backscatter signal at the antenna [7].

Unlike prior backscatter systems, however, we use pulse width modulation (PWM) techniques to convert the analog output of the camera sensor into the digital domain. At a high level, PWM modulation is a single bit analog to digital converter where the analog input voltage translates to the timing information of a digital waveform. Specifically, the output of a PWM modulation is a square wave where the duty cycle of the output square wave is proportional to the analog voltage of the input signal. As we show next, one can implement a PWM converter with passive RC components and a comparator, thereby consuming very low power.

Fig. 4 shows the architecture of our PWM converter. The input is a square wave with amplitude $A$ and operating at frequency $f$, which is determined by the frame rate and resolution of the camera. The square wave is first low pass filtered by an *RC* network to generate an exponential waveform approximating a triangular waveform as shown in the figure. The exponential waveform is compared with the output of the camera sensor to generate the PWM output signal. Below we describe this operation in more detail.

First, let us understand the generation and modeling of the exponential waveform. During the charging process, corresponding to the high logic of the square wave, the capacitor $C$ starts from an initial voltage $V_{min}$. It charges with a time constant $\tau = (R_1 || R_2) C$ to a maximum voltage, $V_{max}$, where $V_{max} \leq V_0 = A \frac{R_2}{R_1 + R_2}$. During the discharging phase when the square wave is at low logic, the capacitor discharges from $V_{max}$ with the same time constant $\tau$. The voltage at the capacitor during the charging and discharging operation can be mathematically written as,

$$V_{charging}(t) = (V_0 - V_{min})(1 - e^{\frac{-t}{\tau}}) + V_{min}$$
$$V_{discharging}(t) = V_{max} e^{\frac{-t}{\tau}}$$

We use a 50% duty cycled square wave as the input. This implies that the charging time $T_1$ and discharging time $T_2$ are equal to half the time period of the square wave, i.e., $T_1 = T_2 = \frac{1}{2f}$. We can use this information to compute the minimum and maximum voltages of the PWM operation as,

$$V_{max} = V_0 \left(1 - e^{\frac{-1}{2f\tau}}\right) + V_{min} e^{\frac{-1}{2f\tau}}$$
$$V_{min} = V_{max} e^{\frac{-1}{2f\tau}}$$

The minimum and maximum voltages in the PWM converter are set by picking the appropriate values for the resistors and capacitors. Specifically, we choose the minimum and maximum voltages to ensure that the camera pixel output $P$ is always within these voltage limits. The exponential waveform is then compared to the output of the camera to generate a PWM digital output. Using the above equations, we can express the duty cycle of the pulse width modulated digital output, $D = pwm(P)$, for each camera pixel, $P$, as,

$$pwm(P) = 0.5 + f\tau \ln\left(\frac{V_0 - P}{V_0 - V_{min}} \times \frac{V_{max}}{P}\right)$$

Hence, by using PWM technique, we convert the analog output of the camera into a digital output corresponding to an open and short impedance state at the antenna, which in turn maximizes radar cross section. This is because the backscatter switch only takes two different values and is in contrast to existing analog backscatter designs, where the backscatter switch takes a range of impedance values.

We note two key points about our design.

1) Traditional backscatter communication systems suffer from self-interference. At a high level, the access point operates as a full-duplex radio and the receiver, in addition to receiving the backscatter signal, also receives a strong interference from the transmitter. Since the sensor data is centered at the carrier frequency, the receiver cannot decode the backscattered data in the presence of a strong in-band interferer. Existing backscatter communication systems such as RFID and Passive Wi-Fi get around this problem by using sub-carrier modulation. These systems use a sub-carrier to shift the signal from the carrier, to a frequency offset $\Delta f$ from the carrier frequency. The receiver can then decode the backscattered signal by filtering the out of band interference.

Our PWM based design can be integrated with subcarrier modulation. Specifically, since our PWM technique converts the camera output into the digital domain, we can implement sub-carrier modulation using digital components. Unlike in the analog domain where sub-carrier modulation is a mixing operation, which requires power hungry analog multipliers, in the digital domain, multiplication operation can be implemented with a simple XOR gate. The sub-carrier can be approximated by a square wave operating at $\Delta f$ frequency and we input the sub-carrier and the PWM output to an XOR gate to up convert the PWM signal to a frequency offset $\Delta f$. Sub-carrier modulation addresses the problem of self-interference at the access point and as a result the PWM backscattering wireless camera can now operate with high SNR and achieve large operating ranges. We show in §4.2 that our PWM backscatter wireless camera can operate upto 150 feet. This is 5x improvement over the lower bandwidth analog backscatter wireless microphone [28].

2) We eliminate the need for amplifiers and AGC in the PWM backscatter redesign of the wireless camera. Traditional systems need amplifiers and AGC to amplify and adjust the sensor output to be within the dynamic range of the ADC. In our design, the PWM converter functions as a one bit analog to digital converter and converts the analog voltage into time domain. Our amplification is performed by the triangular wave that can be generated in a low power manner using passive components and a comparator. Hence, the PWM backscatter design eliminates the need for dedicated power hungry amplifiers and AGC.

## 4. EVALUATION

We evaluate various aspects of our wireless camera system. We start by characterizing the PWM backscatter communication link between the camera and the access point followed by evaluating the quality of the wirelessly received video as a function of separation between the camera and the access point.

### 4.1 PWM Backscatter Characterization

Our wireless camera uses PWM backscatter to transmit analog image sensor information to an access point. We will first evaluate the PWM backscatter technique, which will determine the quality and operating range of the wireless communication link between the camera and the access point.

To isolate the performance of PWM backscatter from the variations due to multipath, our first set of experiments use a wired setup with a variable attenuator and circular. §4.2 describes the end-to-end results over the wireless medium. We set the USRP to transmit a single tone signal at 895 MHz with output power of -8 dBm and connect it to port 1 of the circular using an RF cable. We introduce a variable attenuator using RF cables between the second port of the circular and the antenna terminal of the camera prototype. The variable attenuator mirrors wireless signal propagation loss as the distance between the access point and wireless camera increases. The USRP receiver is connected to the third port of the circular with an RF cable. In this experiment, our goal is to jointly characterize the PWM converter and the PWM backscatter communication link. So, we use 33120A function generator by HP to generate a voltage ramp signal which corresponds to the output voltage of the camera.

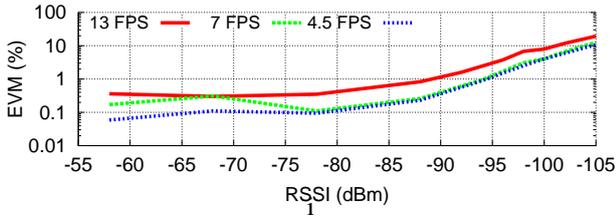

**Figure 5—Sensitivity of PWM Backscatter.** We plot the error vector magnitude (EVM) as the function of receiver signal strength at different frame rates.

We evaluate PWM communication link for different frame rates of the camera. We set the frequency of the voltage ramp signal to 10 Hz. Next, we introduce redundancy in our PWM transmissions to reduce the effective frame rate. Specifically, we transmit 1, 3 and 5 PWM pulses per symbol for PWM communication, which translates to 13, 7 and 4.5 frames per second respectively. We evaluate the PWM communication link for its sensitivity and the effective number of bits available at the receiver.

### 4.1.1 Sensitivity

We start by setting the function generator to transmit a ramp signal from $V_{min}$ to $V_{max}$ which covers the lighting conditions (< 500lux) in indoor environments. We recover the PWM modulated voltage ramp using our access-point and compare the received ramp with the transmitted ramp signal for effective frame rates of 13, 7 and 4.5 frames per second. To evaluate the performance of the PWM backscatter communication link, we compute the difference between the transmitted and received ramp. Specifically, we calculate the error vector magnitude (EVM) as [25], $EVM(\%) = 100 \times \left( \frac{\sum_{i=1}^{N} |S_n - S_0|^2}{\sum_{i=1}^{N} |S_0|^2} \right)$. Here $S_n$ is the received and $S_0$ is the transmitted ramp signal. Fig. 5 shows the *EVM* for different frame rates as a function of the signal strength of the received signal. The plots show the following:

- *EVM* increases as the signal strength of the received signal reduces. This is expected because as the received signal attenuates, the SNR at the receiver decreases which introduces errors in the received ramp signal. The access point can recover PWM backscattered signals with less than 10% error down to -106 dBm.

- As we add redundancy to PWM transmissions and decrease the effective frame rate, the *EVM* also decreases. Redundant transmissions proportionately increase the SNR of the received signal at the receiver and reduce errors in the decoded PWM signal.

- The *EVM* of the received PWM signal at the access point is always between $0.1 - 10\%$ for signals down to -106 dBm which ensures a quality wireless link between the wireless camera and the access point. This translates to a theoretical operating distance of about 328 feet in line of sight scenarios according to the Friis path loss [22].

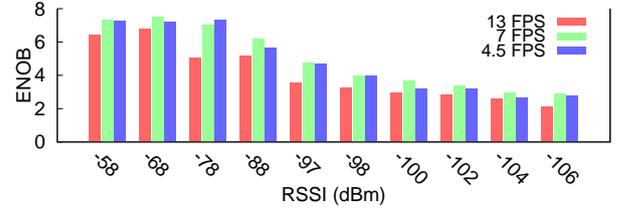

**Figure 6—ENOB at receiver.** We plot the effective number of bits (*ENOB*) at the receiver as a function of the signal strength of the received PWM signal.

### 4.1.2 ENOB at the receiver

PWM backscatter design eliminates the LNA, AGC and ADC at the wireless camera and delegates these power hungry component to the access point. However, the signal at the access point undergoes attenuation during wireless propagation before it is digitized by the ADC at the access point. As a result the SNR at the receiver and the effective number of bits (*ENOB*) or bit resolution is a function of the signal strength at the receiver. The IEEE standard 1057 defines the *ENOB* of an input signal at the ADC as, $ENOB = \log_2 \left( \frac{V_{in}^*}{N_{RMS}\sqrt{2}} \right)$, where $V_{in}^*$ defines the full dynamic range of the recovered voltage and $N_{RMS}$ is the Root Mean Square of the ADC noise. We use the same setup as before and transmit a ramp signal from $V_{min}$ to $V_{max}$ and vary the attenuation and compute the *ENOB* at the receiver. Fig. 6 plots the effective number of bits at the receiver as a function of the signal strength of the received signal for different frame rates. The plot shows that the bit resolution of the received signal decreases with decrease in signal strength and increase in frame rate. Increase in frame and signal attenuation both reduce SNR at the receiver which reduces the *ENOB* of the received signal. When the received signal strength is high, the receiver we can decode signals with 7 bits of resolution and at -106 dBm, the receiver decodes with 3-bit resolution. For context, a 1-bit resolution design can be used to wirelessly transmit black and white video to the receiver.

## 4.2 Operational Range

We deploy our wireless camera system in the parking lot of an apartment complex. We use the USRP based access point implementation and set it to transmit 30 dBm into a 6 dBi patch antenna. This is the maximum power permitted by FCC on the 900 MHz ISM band. We vary the distance between the wireless camera prototype and the access point and at each separation, we stream 20 seconds long video from the camera to the access point using PWM backscatter communication. Simultaneously, we also record the output of the camera using a high input impedance National Instrument USB-6361 DAQ as the ground truth. We use the *PSNR* metric commonly used in video applications to compare the video wirelessly streamed to the access point using PWM backscatter to the ground truth video recorded at the camera using a NI DAQ. *PSNR* computes the difference between the ground truth and wirelessly received video. As a rule of thumb, *PSNR* > 30 *dB* represents negligible loss and *PSNR*

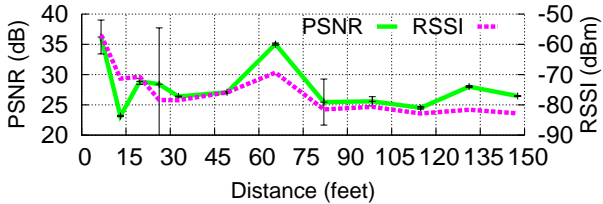

(a) Low Lighting (50-70 lux)

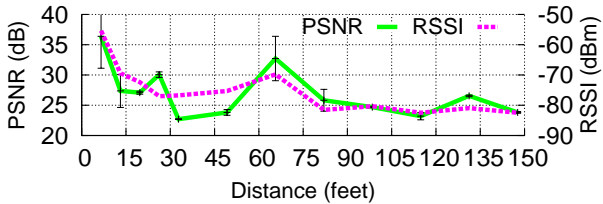

(a) Medium Lighting (100-200 lux)

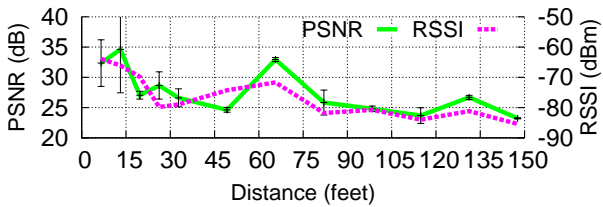

(a) Good Lighting (400-500 lux)

**Figure 7—Operational Range.** We show the *PSNR* of the video wirelessly streamed from the PWM backscatter camera as a function of the distance between the camera and the access point.

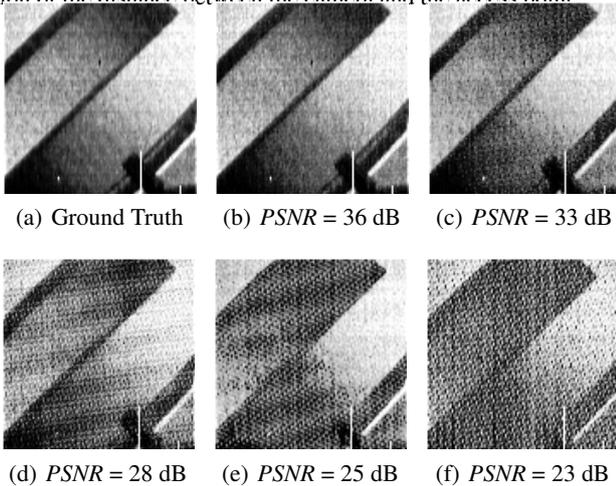

(a) Ground Truth  (b) *PSNR* = 36 dB  (c) *PSNR* = 33 dB

(d) *PSNR* = 28 dB  (e) *PSNR* = 25 dB  (f) *PSNR* = 23 dB

**Figure 8—Sample video frames for different *PSNR*.** We show a video frame for ground truth recorded using the NI DAQ and for a range of *PSNR* values, captured wirelessly.

of above 25 dB shows an acceptable frame quality comparing to the ground truth. Fig 8 shows a snapshot of a video frame of the gray scale ramp image [3] for a range of *PSNR* values.

We measure the performance of the wireless camera for three different lighting conditions at a frame rate of 7 FPS. Fig. 7 plots the *PSNR* of the received video at the access point as a function of the separation between the access point and the wireless camera. The plots show the following,

• Under all lighting conditions, the wireless camera can stream video with an average *PSNR* greater than 24 dB

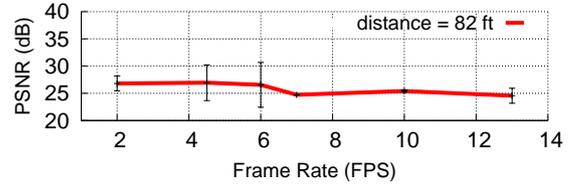

**Figure 9—Adding redundancy to improve *PSNR*.** We reduce frame rate by adding redundancy to PWM transmissions. We plot the *PSNR* of the video as a function of the effective frame rate.

to an access point up to a distance of 150 feet. Beyond 150 feet, the access point cannot reliably decode the sync pulses which limit the operating range of our wireless camera system.

• The *PSNR* of the wirelessly streamed video decreases as the distance between the access point and the camera increases. This is because as the distance increases, the wireless signal undergoes attenuation which reduces the SNR at the receiver and reduces the quality of the received video.

• For most part, the *PSNR* of the received video decreases with increasing distance. However, at certain locations there is variation in the *PSNR* of the received video which can be attributed to wireless multi-path. We also plot the signal strength of the received signal at each location which is correlated with the variations in *PSNR*.

### 4.3 Redundancy versus Video Quality

We showed in §4.1.1 and §4.1.2 that if we reduce the frame rate of the camera by adding redundancy to PWM transmissions, we can improve the SNR, *EVM* and *ENOB* at the receiver. Here we evaluate how these improvements translate to quality of the wireless streamed video. We place the wireless camera at a fixed distance of 82 feet from the access point and vary the frame rate of the camera from 13 FPS down to 2 FPS. Fig. 9 shows that *PSNR* of the received video as a function of the frame rate. We can see that at 13 FPS the video has a *PSNR* of just below 25 dB, the threshold above which the received frame has an acceptable quality comparing to the ground truth. However, if we decrease the frame rate, we can improve the *PSNR* by 3 dB, pushing it over the acceptable threshold at this distance. In deployments where the camera is operating at high frame rate and at the threshold of acceptable *PSNR*, adding redundancy and reducing the effective frame rate gives us the flexibility to improve the performance of the wireless camera.

## 5. PROOF-OF-CONCEPT APPLICATIONS

We demonstrate two proof-of-concept applications.

### 5.1 Micro-Robots

There has been significant recent interest in developing miniature robot such as Kilobots [24], Robo-bees [8], NanoWalkers [30] and Hexapod Robot [20]. The key vision here is that a large number of micro robots can collectively work together and efficiently perform key tasks such as inspections, rescue, monitoring, component assembly as well as artificially pollinate crops [8]. However, power is a key challenge for these micro-robots. Due to payload restrictions, they are constraint to tiny batteries that limit the capabilities and functionality of these robots. For example, it was previously unfeasible to stream video from these robots without significantly limiting the battery life.

To demonstrate the viability of augmenting a micro-robot with our streaming camera technology, we use the Elisa-3 robot from GCtronic [2]. It is a 50 mm diameter robot with two DC motors and is powered using two 3.7 V 130 mAh batteries and is designed to support more weight than that of our camera. During active operation, the robot consumes about 300 mW of power. Streaming digital video consumes about 100s of mW of power for operating ADC and digital communication which would cut the operating time of the robot by at least half. On the other hand, our COTS streaming prototype consumes a maximum of 2.36 mW and will have a negligible impact on the battery life of the robot.

To evaluate the performance of the video streaming camera from a navigating robot, we place our COTS prototype on top of the Elisa-3 robot as shown in Fig. 10(a). We allow the robot to randomly move in a $10 \times 10$ ft$^2$ space and stream 1.5 minutes long videos to the access point located 15 feet away from the robot. Fig. 10(b) shows the CDF of the *PSNR* of the video streamed from robot. The plot shows that the median *PSNR* was above 30 dB while the minimum was 20 dB. This demonstrates that power constrained micro-robots can continuously stream video using our PWM backscatter wireless camera with almost no power overhead.

## 5.2 Security Cameras and Smart Home

Wireless cameras are increasingly popular in security and smart home applications. Unfortunately, these cameras are required to be plugged-in to a power outlet or operate for a limited time on a battery. We show that our ultra-low power wireless camera can be used for security and smart home applications. Specifically, we demonstrate that the quality of the video streamed from our COTS implementation is sufficient for detecting human faces. Such a system can be used to detect human occupancy, grant access (such as Ring [5]) or set off an alarm in case of an intruder. To evaluate the system, we place the wireless camera at five different distances ranging from 16 to 100 feet from the access point. We ask ten users to walk around and perform gestures within 5 feet of the camera. We stream a 2 minute video at each location at 7 FPS. We use the MATLAB implementation of Viola-Jones algorithm to analyze the approximately four thousand video frames. Fig. 11 shows the accuracy of face detection as a function of the *PSNR* of the wirelessly recorded video. The plots show that as the quality (*PSNR*) of the video improves, the accuracy of face detection also increases. We can accurately detect up to 95% of human faces when the *PSNR* of the wirelessly streamed video is greater than 30 dB.

## 6. RELATED WORK

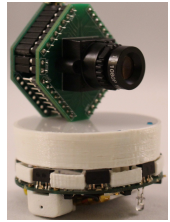 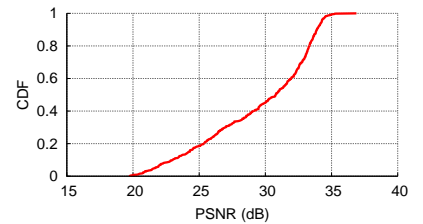

(a) Elisa-3 robot with COTS camera prototype  (b) CDF of PSNR

**Figure 10—Micro-robot.** We show a CDF of the *PSNR* of the video streamed from a micro-robot. Our approach has negligible effect on the battery life of the robot.

Prior work falls in two different categories.

*Backscatter Communication.* One of the early examples of analog backscatter is the gift that was given to the US embassy in Moscow that included a passive listening device. This spy device is a sound-modulated resonant cavity and the voice that moves its diaphragm modulates its resonance frequency [4]. A more recent example of using analog backscatter is a microphone enabled battery-free tag that amplitude modulates its antenna impedance with the output of the microphone [26,28]. In contrast, we design the first video backscatter system using pulse width modulation. Further, prior microphone designs have a low data-rate comparing to video streaming. Our camera at 13 frames per second transmits about $163K$ pixels per second whereas in a microphone case only a few kilo-samples of audio transmission is enough to recover the voice fully. In addition, our camera operates on more than four times the range [28] can operate.

Ekhonet [31] optimizes the computational blocks between the sensor and the backscatter module to reduce the power consumption of backscatter-based wireless sensors. Our design builds on this work but differs from it in multiple ways: 1) prior work still uses ADCs and amplifiers on the cameras to transform the pixels into the digital domain and hence cannot achieve ultra-low-power video streaming. 2) we present a pulse width modulation technique that works with backscatter communication. This allows us to transmit the analog pixels that are generated by the camera through the backscatter switch.

Finally recent work on Wi-Fi and TV based backscatter systems [10–12, 23] can achieve mega bits per second of communication speed using backscatter technique. Integrating these designs with our video backscatter approach is a worthwhile future research direction.

*Low-Power Cameras.* In the past few years, researchers have designed battery-free cameras. [19] introduces a self-powered camera that can switch its photo diodes between energy harvesting and photo capture mode. Despite being self-powered, these cameras do not have wireless data transmission capabilities. [14–16, 18, 27] show that using off-the-shelf camera modules, one can build battery-free wireless cameras that will capture still images using the energy they have harvested from RF waves including Wi-Fi and 900 MHz transmissions. Despite their ability to transmit its data wirelessly, they are heavily duty cycled and cannot stream video. In particular, these designs can send a new frame at most every ten seconds when it is very close to the RF power source (within about a foot) and once every few tens of minutes at more practical distances [15, 27].

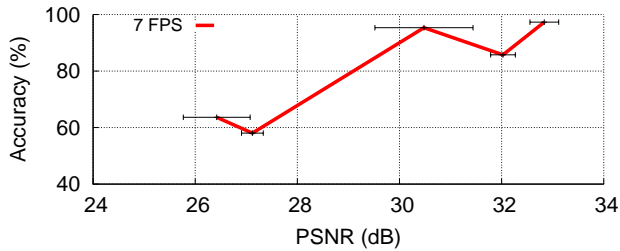

**Figure 11—Face detection.** We show the accuracy of face detection on the video streamed from our wireless camera.

Recent work [13] also addresses the problem that conventional image sensors' power consumption does not scale as their resolution and frame rate. In particular, the authors propose to change the camera input clock as well as aggressively switch the camera to standby mode, based on the desired image quality. This work is however complimentary to our design since streaming video requires addressing the power consumption of multiple components including camera and communication. Our work jointly integrates all these components to achieve the first ultra-low-power video streaming design.

Finally, Glimpse [17] showed that a regular camera on wearable devices burn more than 1200 mW, which limits the operation time of the camera to less than two hours on a wearable device. Glimpse [17] also shows that one can design a low-power gating wearable vision system that only looks for certain events occurrence in the field of view, and turns on the primary imaging pipeline when those events happened. The power and bandwidth saving using Glimpse system is obviously limited to the application and does not deal with communication. In contrast, we present the first ultra-low-power video streaming application by jointly optimizing backscatter communication and camera design and eliminating the traditional power-consuming interfaces such as ADCs and amplifiers between the two modules.